\documentclass[onecolumn]{revtex4}
\usepackage{amsmath}
\usepackage{amsfonts,color}
\usepackage{amssymb,float}
\usepackage{mathtools}
\usepackage[utf8]{inputenc}
\usepackage{url}
\usepackage{graphicx}
\usepackage{refstyle}
\usepackage{wasysym}
\usepackage{accents}
\usepackage{graphicx}
\usepackage{color}
\usepackage[dvipsnames]{xcolor}
\usepackage{hyperref}

\hypersetup{
    colorlinks=true,
    linkcolor=blue,
    filecolor=magenta,      
    citecolor=red
}

\makeatletter
\long\def\dddddot#1{%
  {\mathop {#1}\limits ^{\vbox to-1.4\ex@ {\kern -\tw@ \ex@ \hbox {\normalfont .....}\vss }}}%
}
\long\def\multidots#1#2{%
  \count@=0
  {{\mathop {#2}\limits ^{\vbox to-1.4\ex@ {\kern -\tw@ \ex@ \hbox {\normalfont %
  \loop%
  \ifnum#1>\count@%
  .%
  \advance\count@ by1%
  \repeat%
  }\vss }}}}%
}
\makeatother

\newcommand{\udt}[3]{#1^{#2}_{\phantom{#2}#3}}

\newcommand{\dut}[3]{#1_{#2}^{\phantom{#2}#3}}

\newcommand{\lc}[1]{\accentset{\circ}{#1}}

\begin{document}

\title{Dynamical complexity of the Teleparallel gravity cosmology}

\author{Geovanny A. Rave-Franco}
\email{geovanny.rave@ciencias.unam.mx}
\affiliation{Instituto de Ciencias Nucleares, Universidad Nacional Aut\'{o}noma de M\'{e}xico, 
Circuito Exterior C.U., A.P. 70-543, M\'exico D.F. 04510, M\'{e}xico.}

\author{Celia Escamilla-Rivera}
\email{celia.escamilla@nucleares.unam.mx}
\affiliation{Instituto de Ciencias Nucleares, Universidad Nacional Aut\'{o}noma de M\'{e}xico, 
Circuito Exterior C.U., A.P. 70-543, M\'exico D.F. 04510, M\'{e}xico.}

\author{Jackson Levi Said}
\email{jackson.said@um.edu.mt}
\affiliation{Institute of Space Sciences and Astronomy, University of Malta, Malta, MSD 2080}
\affiliation{Department of Physics, University of Malta, Malta}

\date{\today}

\begin{abstract}
The exploration of teleparallel gravity has been done from a dynamical systems point of view in order to be tested against the cosmological evolution currently observed. So far, the proposed autonomous systems have been restrictive over a constant dynamical variable, which contains information related to the dynamics on the $H_0$ value. It is therefore that in this paper we consider a generalization of the dynamical system by imposing a nonconstant degree of freedom over it which allows us to rewrite a generic autonomous dynamical analysis. We describe the treatment of our nonlinear autonomous system by studying the hyperbolic critical points and discuss an interesting phenomenological feature in regards to $H_0$: the possibility to obtain a best-fit value for this parameter in a cosmologically viable $f(T,B)$ model, a mixed power law. This result allows us to present a generic scenario in which it is possible to fix constraints to solve the $H_0$ tension at late times where its linearized solutions are considered.
\end{abstract}

\maketitle

\section{Introduction}
$\Lambda$CDM cosmology offers a very successful model structure in which to study galactic dynamics through the prism of cold dark matter \cite{Baudis:2016qwx}, whereas cosmological-scale physics is dominated by a cosmological constant \cite{weinberg2008cosmology}. Together with an inflationary epoch \cite{Perenon:2015sla}, $\Lambda$CDM cosmology largely reproduces the correct observations measured at various scales of the Universe. The necessity of adding both dark matter and dark energy stems from the lack of suitable predictions from general relativity (GR) which is the gravitational foundation of $\Lambda$CDM. Despite the success of $\Lambda$CDM as a cosmological model, some crucial observations have started to show important tensions between the model and observational evidence, which are apart from the well-known theoretical problems internal to the theory \cite{Weinberg:1988cp,Clifton:2011jh}. The prominent contention in recent years has been the $H_0$ tension in which local and early Universe observations predict different values of the $H_0$ expansion parameter \cite{DiValentino:2020zio}. Another growing tension in recent cosmological data is related to the growth of large-scale structure in the $f\sigma_8$ parameter \cite{DiValentino:2020vvd}. Also, a more novel tension has been suggested wherein cosmic birefringence would be present in the \textit{Planck} Collaboration 2018 data release \cite{Minami:2020odp} which would pose an even more serious problem for $\Lambda$CDM. \medskip

The growing observational tensions motivate us to explore the possible landscape of gravitational theories beyond GR which can each be distinguished through Lovelock's theorem \cite{Lovelock:1971yv} which provides a clear set of criteria through which to introduce different forms of gravity. One of these potential theories of gravity is that of teleparallel gravity (TG) where the curvature produced by the Levi-Civita connection is replaced by the torsion associated with the teleparallel connection \cite{Weitzenbock1923,Aldrovandi:2013wha}. This novel foundation of gravity does not depend on GR but can produce a teleparallel equivalent to general relativity (TEGR) which is dynamically equivalent to GR in terms of its field equations \cite{Krssak:2018ywd,Cai:2015emx} but that originates from a distinct action in which the Ricci scalar Lagrangian is replaced by a torsion scalar \cite{Farrugia:2016qqe,Farrugia:2016xcw}. The core difference between these two Lagrangians is that the Ricci scalar also contains total divergence terms which has the impact that the teleparallel analogue of the Lovelock theorem will produce a much larger class of gravitational models which retain generically second-order derivatives in their field equations \cite{Gonzalez:2015sha,Gonzalez:2019tky}. TEGR has also been shown to have a number of other attractive features such as its likeness to a Yang-mills theory \cite{Aldrovandi:2013wha} giving it a strong similarity to a particle physics theory, its potential definition of a gravitational energy-momentum tensor \cite{Blixt:2018znp,Blixt:2019mkt}, as well as it not requiring a Gibbons-Hawking-York boundary term giving TEGR a well defined Hamiltonian description, among other properties. Observationally and more generally, TG also has a number of attractive features as well such as Refs.\cite{Briffa:2020qli,LeviSaid:2020mbb,Finch:2018gkh} (which refer to observational works in $f(T)$ theories of gravity) and Ref.\cite{Bahamonde:2020bbc} (which explored solar system tests in $f(T,B)$ theories of gravity). \medskip

In the same vein as $f(\lc{R})$ gravity \cite{DeFelice:2010aj,Capozziello:2011et} (over-circles denote quantities determined using the Levi-Civita connection), TEGR can be directly generalized to $f(T)$ gravity \cite{Ferraro:2006jd,Ferraro:2008ey,Bengochea:2008gz,Linder:2010py,Chen:2010va,Bahamonde:2019zea,Ualikhanova:2019ygl}, which is a second-order gravitational theory that has shown several potentially positive observational predictions \cite{Cai:2015emx,Nesseris:2013jea,Farrugia:2016qqe,Finch:2018gkh,Farrugia:2016xcw,Iorio:2012cm,Ruggiero:2015oka,Deng:2018ncg,Yan:2019gbw,LeviSaid:2020mbb,Paliathanasis:2017htk}. However, to fully embrace the generalizations offered by TG and relate this generalization to $f(\lc{R})$ gravity, we must also consider the inclusion of the boundary (or total divergence) term, $B$, through $f(T,B)$ gravity \cite{Bahamonde:2015zma,Capozziello:2018qcp,Bahamonde:2016grb,Paliathanasis:2017flf,Farrugia:2018gyz,Bahamonde:2016cul,Bahamonde:2016cul,Wright:2016ayu}. This relation becomes an equivalence when the arguments are combined in a particular prescribed way, namely when the limit $f(T,B) = f(-T+B) = f(\lc{R})$ is selected. Similar to $f(T)$ gravity, there have been a number of studies in this extension to TEGR \cite{Farrugia:2020fcu,Capozziello:2019msc,Farrugia:2018gyz,Bahamonde:2015zma,Paliathanasis:2017flf,Bahamonde:2016grb,Bahamonde:2016cul,Bahamonde:2015zma,Escamilla-Rivera:2019ulu,Franco:2020lxx} which have shown a number of promising results for the theory.

TG also has a number of other extensions to TEGR such as $f(T,T_G)$ gravity where the TG analogue of the Gauss-Bonnet scalar is incorporated into the gravitational model \cite{Kofinas:2014daa,Capozziello:2016eaz,delaCruz-Dombriz:2018nvt,delaCruz-Dombriz:2017lvj}. Another recent proposal of TG was proposed in Ref.\cite{Bahamonde:2019shr} where a new term in addition to the standard Horndeski terms was introduced which was later found to allow for a number of Horndeski models to be revived despite the recent speed of light constraint on the propagation of gravitational waves \cite{Bahamonde:2019ipm}. This new realization of Horndeski gravity has also been explored through the parametrized post-Newtonian (PPN) formalism \cite{Bahamonde:2020cfv} where it was found that most models largely survive current PPN constraints.

Dynamical systems offers a vital approach in which the background equations of cosmological models can be probed against cosmological observations \cite{Bahamonde:2017ize}. In this work, we probe $f(T,B)$ gravity in the context of a homogeneous and isotropic universe using the Friedmann–Lema\^{i}tre–Robertson–Walker (FLRW) metric. This approach has been used to explore higher-order modified TG models in Ref.\cite{Karpathopoulos:2017arc} where a number of important reconstructions were investigated together with their dynamical systems evolution, and where they also compared their results with supernova type Ia data. In Ref.\cite{Bahamonde:2016grb}, $f(T,B)$ gravity was directly studied in the context of Noether's theorem where several reconstruction approaches were explored along with some stability conditions. Another important work where Noether's theorem was used to determine new solutions is Ref.\cite{Capozziello:2014bna}, some of which have shown promise in terms of producing cosmologically viable models \cite{Bahamonde:2016grb}. $f(T,B)$ gravity has also been investigated in terms of its thermodynamics in Ref.\cite{Bahamonde:2016cul}, where the matter perturbation equation was derived. Other reconstructions \cite{Pourbagher:2019zhq,Zubair:2018wyy} have been developed in this class of theories of gravity where energy condition information was also provided.

In Ref.\cite{Escamilla-Rivera:2019ulu} a number of literature $f(T,B)$ gravity models were fitted using cosmic chronometers, supernova and baryonic acoustic oscillation data which resulted in model parameter constraints. In Ref.\cite{Franco:2020lxx}, the viable models were explored in terms of their dynamical systems. One of these realizations was the mixed power-law model where the torsion scalar and boundary term are coupled with independent indices. This is interesting because it is one of the core studied models in the literature that is novel from $f(\lc{R})$ gravity, i.e. that cannot be produced by any choice of this free function. However, due to the complexity of the ensuing dynamical system, one of the dynamical variables had been set to be a constant which overly constrained the ensuing system. In this work, we find an approach in which this assumption does not need to be taken and the fully autonomous dynamical system can be explored for this model. The paper is structured as follows: in Sec.~\ref{sec: cosmology_fTB}, we develop the technical details of $f(T,B)$ gravity and provide the background cosmology equations. The dynamical system is then defined in Sec.~\ref{sec:dynamicalsec} for the mixed power-law model of $f(T,B)$ gravity. It is in this section that we generalize the dynamical variable from being a constant. In Sec.~\ref{sec:linearised}, we describe the dynamical systems treatment in further detail and expose the hyperbolic critical points of the system. The possibility of reducing the $H_0$ tension is discussed in Sec.~\ref{sec:h0tension} where we obtain a best-fit value for this observational parameter. Finally, in Sec.~\ref{sec:conclusions}, we summarize our core results and provide a discussion of these results.

\section{\texorpdfstring{$f(T,B)$}{f(T,B)} cosmology}
\label{sec: cosmology_fTB}

TG is built on the exchange of the curvature associated with the Levi-Civita connection $\mathring{\Gamma}^{\sigma}_{\mu\nu}$ (we use over-circles to denote quantities calculated with the Levi-Civita connection throughout) with the torsion linked to the teleparallel connection $\Gamma^{\sigma}_{\mu\nu}$ \cite{Cai:2015emx,Krssak:2018ywd}. GR and its modifications are principally built from scalars based on the Riemann tensor \cite{Clifton:2011jh}. However, the Riemann tensor identically vanishes when the teleparallel connection is used since it is curvatureless and satisfies metricity.

The construction of the teleparallel connection is formulated by building the metric tensor $g_{\mu\nu}$ from tetrad fields $\udt{e}{A}{\mu}$ (and their inverses $\dut{E}{A}{\mu}$) \cite{Aldrovandi:2013wha}. In this setting, latin indices refer to local Minkowski space coordinates while greek indices refer to the general manifold, where the tetrad components solder both spaces together and provide a direct approach to transform indices between these manifolds. These concepts come together in the relations that produce the metric tensor from the tetrad fields which take the form
\begin{align}\label{metric_tetrad_eq}
    g_{\mu\nu}=\udt{e}{A}{\mu}\udt{e}{B}{\nu}\eta_{AB}\,,& &\eta_{AB}=\dut{E}{A}{\mu}\dut{E}{B}{\nu}g_{\mu\nu}\,,
\end{align}
where the tetrads satisfy the orthogonality conditions
\begin{align}
    \udt{e}{A}{\mu}\dut{E}{B}{\mu}=\delta^B_A\,,& &\udt{e}{A}{\mu}\dut{E}{A}{\nu}=\delta^{\nu}_{\mu}\,,
\end{align}
for internal consistency. Then, the teleparallel connection can be defined as \cite{Weitzenbock1923}
\begin{equation}
    \Gamma^{\sigma}_{\nu\mu} := \dut{E}{A}{\sigma}\partial_{\mu}\udt{e}{A}{\nu} + \dut{E}{A}{\sigma}\udt{\omega}{A}{B\mu}\udt{e}{B}{\nu}\,,
\end{equation}
where $\udt{\omega}{A}{B\mu}$ represents the flat (or inertial) spin connection \cite{Golovnev:2017dox,Hohmann:2018rwf,Bejarano:2019fii}. The spin connection appears in the connection to sustain the general covariance of the ensuing theories \cite{Krssak:2015oua}. The components of the spin connection are totally inertial and represent the local Lorentz transformations (LLTs) \cite{Aldrovandi:2013wha,Hehl:1994ue}. Naturally, one may choose a Lorentz frame in which the spin connection components vanish which is called the Weitzenb\"{o}ck gauge. However, one must be careful when applying this gauge in order to not lose LLT invariance of the equations of motion.

Taking the same rationale as GR where scalars are built from the Riemann tensor, TG theories are constructed using contractions of the torsion tensor defined as \cite{Krssak:2018ywd,Cai:2015emx}
\begin{equation}
    \udt{T}{\sigma}{\mu\nu} :=- 2\Gamma^{\sigma}_{\left[\mu\nu\right]}\,,
\end{equation}
where square brackets denote the usual antisymmetric operator. The torsion tensor represents the field strength of TG, and can produce a number of other measures of torsion \cite{Aldrovandi:2013wha}. There exists a particular choice of quadratic contractions of the torsion tensor called the torsion scalar defined as
\begin{equation}\label{torsion_scalar_def}
	T := \frac{1}{4}\udt{T}{\sigma}{\mu\nu}\dut{T}{\sigma}{\mu\nu} + \frac{1}{2} \udt{T}{\sigma}{\mu\nu} \udt{T}{\nu\mu}{\sigma} - \udt{T}{\sigma}{\mu\sigma}\udt{T}{\nu\mu}{\nu}\,,
\end{equation}
which is the TG analogue of the standard gravity Ricci scalar $\lc{R}$. The introduction of the torsion scalar is necessary because the Ricci scalar identically vanishes when the teleparallel connection is assumed, i.e. $R=0$.

Calculating the Ricci scalar in both connections leads to a revealing relation from which the torsion scalar emerges, namely \cite{Hayashi:1979qx,Hehl:1976kj}
\begin{equation}\label{Ricci_torsion_equiv}
    R=\lc{R} + T - B = 0\,,
\end{equation}
where $B:=\frac{2}{e}\partial_{\mu}\left(e T^{\mu}\right) = \mathring{\nabla}_{\mu}\left(T^{\mu}\right)$ is a boundary term, and $e = {\rm det}\left(\udt{e}{A}{\mu}\right)=\sqrt{-g}$ is the tetrad determinant. Hence, a TEGR action can be written as
\begin{equation}
    \mathcal{S}_{\rm TEGR} = -\frac{1}{2\kappa^2}\int d^4 x\, eT + \int d^4 x\, e\mathcal{L}_m\,,
\end{equation}
where $\kappa^2=8\pi G$ and $\mathcal{L}_m$ is the regular matter Lagrangian, which are guaranteed to reproduce the Einstein equations.

The boundary term plays no role in the ensuing TEGR field equations as one would expect. However, in standard gravity GR the boundary term is the reason why many modified theories of gravity turn out to be fourth order in their field equations, such as in $f(\lc{R})$ gravity \cite{Sotiriou:2008rp,Capozziello:2011et}. Thus, in TG, we decouple the second- and fourth-order contributions to the field equations. In the context of the central role that the torsion scalar and the boundary term play in TG and its relation to GR, the generalization of TEGR to $f(T,B)$ gravity has gained significant momentum in recent years \cite{Franco2020,Bahamonde:2015zma,Capozziello:2018qcp,Bahamonde:2016grb,Paliathanasis:2017flf,Farrugia:2018gyz,Bahamonde:2016cul,Bahamonde:2016cul,Wright:2016ayu}. The theory has an associated action
\begin{equation}\label{f_T_B_action}
    \mathcal{S}_{f(T,B)} = \frac{1}{2\kappa^2}\int d^4 x\, e f(T,B) + \int d^4 x\, e\mathcal{L}_m\,.
\end{equation}
where it is important to note that $f(T,B)=f(-T+B)=f(\lc{R})$, but more generally $f(T,B) \neq f(\lc{R})$.

In the present work, we work in a flat FLRW universe where the tetrad field
\begin{equation}\label{flrw_tetrad}
    \udt{e}{A}{\mu}={\rm diag}(1,a(t),a(t),a(t))\,,
\end{equation}
turns out to be compatible with the Weitzenb\"{o}ck gauge in $f(T,B)$ gravity \cite{Bahamonde:2016cul,Tamanini:2012hg,Caruana:2020szx}, and where $a(t)$ is the scale factor. Through Eq.(\ref{metric_tetrad_eq}), this leads directly to the regular FLRW metric
\begin{equation}
    ds^2=dt^2-a(t)^2(dx^2+dy^2+dz^2)\,.
\end{equation}

Through the definitions of the torsion scalar and boundary term, it follows that for a flat FLRW scenario
\begin{equation}
	T = 6H^2\,, \quad B = 6(3H^2+\dot{H})\,,
\end{equation}
which together reproduce the Ricci scalar, i.e. $\lc{R}=-T+B = 6(\dot{H} + 2H^2)$. These scalars contribute to produce the $f(T,B)$ gravity Friedmann equations
\begin{align}
	&-3H^2\left(3f_B + 2f_T\right) + 3H\dot{f}_B - 3\dot{H} f_B + \frac{1}{2}f = \kappa^2\rho\label{Friedmann_1}\,, \\
	&-\left(3H^2+\dot{H}\right)\left(3f_B + 2f_T\right) - 2H\dot{f}_T + \ddot{f}_B + \frac{1}{2}f = -\kappa^2 P\label{Friedmann_2}\,,
\end{align}
where overdots refer to derivatives with respect to cosmic time $t$, and where $\rho$ and $P$ respectively represent the energy density and pressure of matter.

In order to better explore the contributions from the modified Lagrangian, we consider $f(T,B)$ gravity as an effective fluid that appears in addition to TEGR through $f(T,B)\rightarrow-T + \tilde{f}(T,B)$. Using this setting, the $f(T,B)$ Friedmann equations can be written as
\begin{align}
    3H^2 &= \kappa^2 \left(\rho+\rho_{\rm eff}\right)\,,\\
    3H^2 + 2\dot{H} &= -\kappa^2\left(P+P_{\rm eff}\right)\,,
\end{align}
where the effective energy density and pressure terms take the form
\begin{align}
    \kappa^2 \rho_{\rm eff} &:= 3H^2\left(3\tilde{f}_B + 2\tilde{f}_T\right) - 3H\dot{\tilde{f}}_B + 3\dot{H}\tilde{f}_B - \frac{1}{2}\tilde{f}\,, \label{eq:friedmann_mod}\\
  \kappa^2 P_{\rm eff} &:= \frac{1}{2}\tilde{f}-\left(3H^2+\dot{H}\right)\left(3\tilde{f}_B + 2\tilde{f}_T\right)-2H\dot{\tilde{f}}_T+\ddot{\tilde{f}}_B\,.
\end{align}

An interesting property of this effective fluid is that it continues to observe continuity equation 
\begin{equation}
    \dot{\rho}_{\rm eff}+3H\left(\rho_{\rm eff}+P_{\rm eff}\right) = 0\,,
\end{equation}
and also defines an effective equation of state (EoS) parameter
\begin{align}\label{EoS_func}
    \omega_{\rm eff} &:= \frac{P_{\rm eff}}{\rho_{\rm eff}}\nonumber\\ 
    &=  -1+\frac{\ddot{\tilde{f}}_B-3H\dot{\tilde{f}}_B-2\dot{H}\tilde{f}_T-2H\dot{\tilde{f}}_T}{3H^2\left(3\tilde{f}_B+2\tilde{f}_T\right)-3H\dot{\tilde{f}}_B+3\dot{H}\tilde{f}_B-\frac{1}{2}\tilde{f}}\,. 
\end{align}
As expected, this limits to the $\Lambda$CDM model when $f(T,B)$ is a constant.

\section{Generic \texorpdfstring{$f(T,B)$}{} cosmological dynamical system}
\label{sec:dynamicalsec}

In our previous work \cite{Franco2020} it was noted that the mixed power-law model
\begin{equation}
    f(T,B) = f_0 B^k T^m\,,
\end{equation}
leads to a degeneration in the dynamical variables, which allows us to study the general case of $\lambda= \frac{\ddot{H}}{H^3}$ and still work with an autonomous dynamical system. In fact, 
\begin{equation}
    X = f_B = k f_0 B^{k-1} T^m = \frac{k}{B}f = \frac{k}{6(3H^2 + \dot{H})}f = \frac{f}{6H^2}\frac{k}{3 +\frac{\dot{H}}{H^2}}=-\frac{Wk}{3+Z}\,, \label{x dependent}
\end{equation}
with $X=f_B$, $W = -\frac{f}{6H^2}$ and $Z = \frac{H'}{H}$, where the prime denotes the derivative with respect to $N = \ln a$. It can be noted that $X$ is not an independent variable, and it depends on $W$ and $Z$, with the restriction $Z \neq -3$. This corresponds to a nonvanishing boundary term $B \neq 0$, since $B=6H^2(3+Z)$.

Following the same reasoning, it is easy to show that $Y=f'_B$ is not an independent variable. Then, using the second Friedmann equation it is possible obtain the derivative $ \frac{d}{dN}\lambda$ in terms of $Z,W,\lambda$ and close the dynamical system. Nevertheless, in doing so, it is possible but cumbersome to actually solve the system but this would require a different approach. From Eq.(\ref{x dependent}) it is required that $Z\neq -3$, where it is then possible to isolate $W$
\begin{equation}
    W = -\frac{X}{k} (Z+3)\,,\label{w dependent}
\end{equation}
then, instead of considering $Z$, $W$ and $\lambda$ as dependent variables, we consider $X$, $Y$ and $Z$ as independent variables and $W$, $\lambda$ as dependent ones. This approach is equivalent to the other one but easier to work with in practical terms. In fact, a direct calculation using Eq.(\ref{w dependent}) and its derivative, shows that $\lambda= \frac{\ddot{H}}{H^3}$ can be rewritten as
\begin{equation}
    \lambda = \frac{1}{1-k}\left(6 k Z+2 (m-1) (Z+3) Z-\frac{Y (Z+3)}{X}+2 Z^2 \right)\,.\label{lambda}
\end{equation}
Therefore, if we consider a universe filled with two fluids $\rho = \rho_{\omega} + \rho_r$ where $\rho_{\omega}$ is a perfect fluid in which the EoS $\omega$ does not necessarily vanish, and $\rho_{r}$ is the EoS for radiation, then the independent dynamical variables are
\begin{equation}
    X \equiv f_B\,, \quad Y = f'_B\,,\quad Z = \frac{H'}{H} = \frac{\dot{H}}{H^2}\,, \quad V \equiv \Omega_r \equiv \frac{\kappa \rho_r}{3H^2}\,.
\end{equation}
In this context, the Friedmann equations can be recast in terms of a dynamical system given through the set of coupled differential equations
\begin{align}
    Z' &= \lambda - 2Z^2\,, \\
    X' &= Y\,, \\
    Y' &= - 3\omega\Omega_{\omega} - V + 3W + (9 + 3Z)X + f_T(6 + 2Z) + 2 f'_T - ZY - 3 - 2Z\,, \\
    V' &= -4V -2ZV\,,
\label{eq:variables_system4}
\end{align}
where $f_T = -mW$, $W$ and $\lambda$ are given by Eqs.(\ref{w dependent}) and Eq.(\ref{lambda}) respectively, and
\begin{equation}
    \Omega_{\omega} = 1 - V - \Omega_{\rm eff}\,,
\end{equation}
where
\begin{equation} \label{density eff}
    \Omega_{\rm eff} = (3+Z)X +2f_T -Y +W\,.
\end{equation}

\noindent The critical point of the dynamical system is
\begin{equation}
    \mathbf{x}_{*} = (Z,X,Y,V) = \left(0,\frac{k}{3 (k+2 m-1)},0,0\right)\,, \label{critical point}
\end{equation}
and the stability matrix evaluated at the critical point is
\begin{equation} \label{stability matrix at x_0}
    \left.\mathcal{M}\right |_{\mathbf{x}_*}=\left(
    \begin{array}{cccc}
     -\frac{6 m}{k-1}-6 & 0 & \frac{9 (k+2 m-1)}{(k-1) k} & 0 \\
     0 & 0 & 1 & 0 \\
     -\frac{2 m}{k-1}+\omega -1 & \frac{9 (\omega +1) (k+2 m-1)}{k} & \frac{6 m}{k-1}-3 \omega  & 3 \omega -1 \\
     0 & 0 & 0 & -4 \\
    \end{array}
    \right)\,.
\end{equation}
The eigenvalues of Eq.(\ref{stability matrix at x_0}) are given by
\begin{align}
\mu_1 &= -4 \,,\\
\mu_2 &= -\frac{3}{2}\left(1 + \frac{\sqrt{\alpha}}{k(k-1)} \right)\,,\\
\mu_3 &= -\frac{3}{2}\left(1 - \frac{\sqrt{\alpha}}{k(k-1)} \right)\,,\\
\mu_4 &= -3(1+\omega)\,,
\end{align}
where $\alpha = (k-1) k \left(9 k^2+24 (k-1) m-17 k+16 m^2+8\right)$. In order to obtain a stable critical point using linear theory \cite{Perko2008,Bahamonde2018}, we have to determine what values of $\omega$, $k$ and $m$ make the real part of the eigenvalues vanish, i.e. make the critical point nonhyperbolic. From $\mu_{4}$, we observe that the value $\omega = -1$ leads us to a nonhyperbolic critical point; then the condition $\omega \neq -1$ is required. \medskip

On the other hand, observe that if $\alpha <0$ then all eigenvalues have nonvanishing real parts, and therefore let us focus on the case $\alpha >0$ that makes all eigenvalues real. We considering each eigenvalue in turn
\begin{itemize}
\item $\mu_2 =0$:
    \begin{enumerate}
        \item \begin{equation}
            0<m<\frac{1}{2}\land (k=1-2 m\lor k=1-m)\,.
        \end{equation}
        \item \begin{equation}
            \frac{1}{2}\leq m<1\land k=1-m\,.
        \end{equation}
    \end{enumerate}
\item $\mu_3 = 0$:
    \begin{enumerate}
        \item \begin{equation}
            m<0\land (k=1-m\lor k=1-2 m)\,.
        \end{equation}
        \item \begin{equation}
            \frac{1}{2}<m\leq 1\land k=1-2 m\,.
        \end{equation}
        \item \begin{equation}
            m>1\land (k=1-2 m\lor k=1-m)\,.
        \end{equation}
    \end{enumerate}
\end{itemize}

Now, let us examine the conditions over $\omega$, $k$ and $m$ which provide a stable critical point. The region of any kind of instability (saddle-like or repulsor-like) is not going to be stated explicitly since, as we shall see later, the critical point is a de Sitter acceleration phase, and then any kind of instability of the critical point is not supported by observations. As stated earlier, the condition that $\alpha \le 0$ gives, without any further condition, a stable critical point and, based on the eigenvalue $\mu_4$. In order to have a stable hyperbolic critical point, the inequality $\omega > -1$ is a necessary condition. Hence any kind of ghost-like fluid generates an inevitable instability. The conditions over $k$ and $m$ such that $\alpha \le 0$ are as follows.
\begin{enumerate}
\item 
    \begin{equation}
        m\leq -\frac{1}{48}\land 0< k< 1\,.
    \end{equation}
\item 
    \begin{align}
        &-\frac{1}{48}<m<0\land \nonumber\\ 
        &\Bigg(0< k\leq \frac{1}{18} (17-24 m)-\frac{1}{18} \sqrt{48 m+1}\lor \frac{1}{18} (17-24 m) \nonumber\\
        & +\frac{1}{18} \sqrt{48 m+1}\leq k< 1\Bigg)\,. \nonumber
    \end{align}
\item 
    \begin{equation}
        m=0\land 0< k\leq \frac{8}{9}\,.
    \end{equation}
\item 
    \begin{align}
        &0<m<\frac{1}{2}\land \nonumber\\
        &\Bigg(0\leq k\leq \frac{1}{18} (17-24 m)-\frac{1}{18} \sqrt{48 m+1}\lor \frac{1}{18} (17-24 m) \nonumber\\
        & +\frac{1}{18} \sqrt{48 m+1}\leq k\leq 1\Bigg)\,. \nonumber
    \end{align}
\item 
    \begin{equation}
        m=\frac{1}{2}\land \frac{5}{9}\leq k<1\,.
    \end{equation}
\item 
    \begin{align}
        &\frac{1}{2}<m<1\land \nonumber\\ 
        &\Bigg(\frac{1}{18} (17-24 m)-\frac{1}{18} \sqrt{48 m+1}\leq k< 0\lor \frac{1}{18} (17-24 m) \nonumber\\
        & +\frac{1}{18} \sqrt{48 m+1}\leq k< 1\Bigg)\,. \nonumber
    \end{align}
\item 
    \begin{equation}
        m=1\land -\frac{7}{9}\leq k< 1\,.
    \end{equation}
\item 
    \begin{align}
        &m>1\land \nonumber\\ 
        &\Bigg(\frac{1}{18} (17-24 m)-\frac{1}{18} \sqrt{48 m+1}\leq k\leq \frac{1}{18} (17-24 m) \nonumber\\
        & +\frac{1}{18} \sqrt{48 m+1}\lor 0< k< 1\Bigg)\,.\nonumber
    \end{align}
\end{enumerate}

Now, the condition $\alpha >0$ leads to a set of real eigenvalues; then to determine the stability of the critical point, we have to analyze the regions such that $\mu_2 <0$ and $\mu_3<0$. These regions are as follows.
\begin{enumerate}
\item 
    \begin{equation}
        m\leq -\frac{1}{48}\land 1-m<k<1-2 m\,.
    \end{equation}
\item 
    \begin{align}
        &-\frac{1}{48}<m\leq 0\land \nonumber\\
        &\Big(\frac{1}{18} (17-24 m)-\frac{1}{18} \sqrt{48 m+1}<k<\frac{1}{18} (17-24 m)+\frac{1}{18} \sqrt{48 m+1}\lor \nonumber \\ 
        &1-m<k<1-2 m\Big)\,.
    \end{align}
\item 
    \begin{align}
        &0<m\leq \frac{1}{2}\land \nonumber\\ 
        &\left(\frac{1}{18} (17-24 m)-\frac{1}{18} \sqrt{48 m+1}<k<1-2 m\lor\right. \nonumber \\ &\left. 1-m<k<\frac{1}{18} (17-24 m)+\frac{1}{18} \sqrt{48 m+1}\right)\,.
    \end{align}
\item 
    \begin{align}
        &\frac{1}{2}<m\leq 1\land \nonumber\\ 
        &\left(1-2 m<k<\frac{1}{18} (17-24 m)-\frac{1}{18} \sqrt{48 m+1}\lor\right. \nonumber \\ 
        &\left. 1-m<k<\frac{1}{18} (17-24 m)+\frac{1}{18} \sqrt{48 m+1}\right)\,.
    \end{align}
\item 
    \begin{align} 
        &m>1\land \nonumber\\ 
        &\left(1-2 m<k<\frac{1}{18} (17-24 m)-\frac{1}{18} \sqrt{48 m+1}\lor\right. \nonumber \\ &\left. \frac{1}{18} (17-24 m)+\frac{1}{18} \sqrt{48 m+1}<k<1-m\right)\,.
    \end{align}
\end{enumerate}
The regions different to the aforementioned values lead to either saddle-like instability or repulsor-like instability. \medskip

\begin{figure}[t]
\begin{center}
\includegraphics[scale=1]{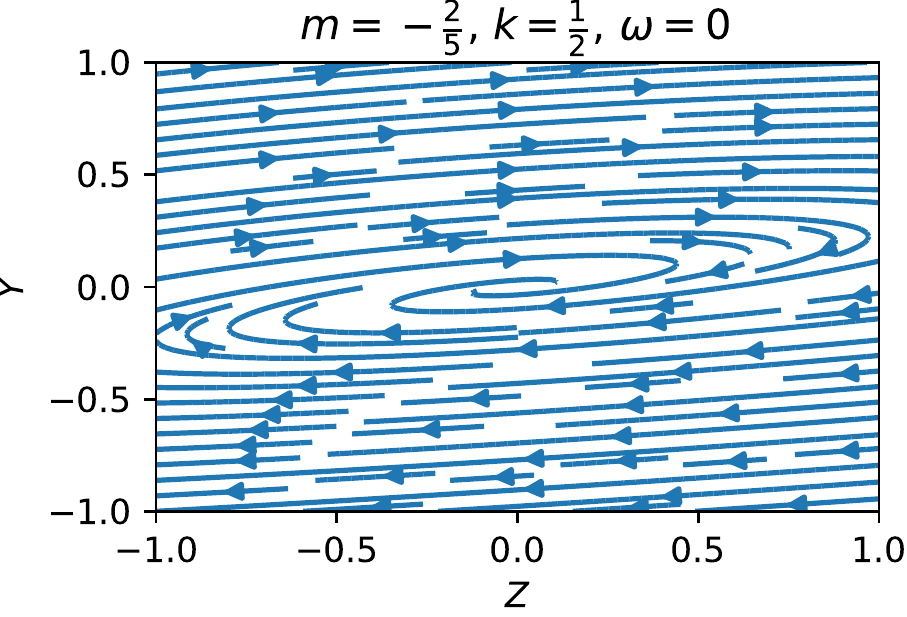}
\caption{Two dimensional phase portrait of the dynamical system Eq.(\ref{eq:variables_system4}) for $m=-2/5$, $k=1/2$ and $\omega = 0$ with $X = \frac{k}{3(-1+k+2m)}$, $V =0$. Here we notice the attractor nature of the critical point of the four-dimensional dynamical system.}
\label{2dportrait}
\end{center}
\end{figure}

While evaluating the density parameter of the effective fluid in Eq.(\ref{density eff}) at the critical point, it is straightforward to find that
\begin{equation}
    \Omega_{\rm eff} = 1\,,
\end{equation}
and $\Omega_r = 0$, $\Omega_{\omega} = 0$. Therefore, the critical point corresponds to an effective-fluid-dominated era, and since $Z = 0$ at the critical point and $Z=\frac{\dot{H}}{H^2}$, then $H = H_0$, which corresponds to a de Sitter universe. Actually, when evaluating the deceleration parameter $q=-\ddot{a}a/\dot{a}^2$ and the jerk parameter $j = \dddot{a}/aH^3$ at the critical point, we obtain
\begin{equation}
    q = -1\,, \quad j =1\,,
\end{equation}
respectively. Here, it is important to point out that this is radically different to our previous result in Ref.\cite{Franco:2020lxx}. It was shown in Ref.\cite{Franco2020} that in the mixed power-law model, when considering $\lambda$ as a constant, the critical points corresponded to a matter-dominated era, therefore, when turning on $\lambda$ to change dynamically, the matter-dominated critical points are shifted to a single de Sitter point. Therefore, the mixed power-law model is able to recover a late-time accelerated universe as an attractor point. In Fig.~\ref{2dportrait} a two-dimensional phase portrait of the dynamical system from (\ref{eq:variables_system4}) is shown.

\section{Linearized \texorpdfstring{$f(T,B)$}{} solutions near the critical point}
\label{sec:linearised}

Consider an autonomous dynamical system of the form
\begin{equation}
    \dot{\mathbf{x}} = \mathbf{f}(\mathbf{x})\,, \label{dynamical}
\end{equation}
where $\mathbf{x}\in \mathbb{R}^n$, $\mathbf{x}_0$ is a critical point of the system and $\mathbf{f} \in C^{r}\left( U_{\mathbf{x}_0}\right)$, $r\ge 1$ with $U_{\mathbf{x}_0}$ being a neighborhood of $\mathbf{x}_0$. Since  $\mathbf{f} \in C^{r}\left( U_{\mathbf{x}_0}\right)$ we could approximate $\mathbf{f}$ in the neighborhood $ U_{\mathbf{x}_0}$ by a Taylor polynomial; then we have that the autonomous dynamical system is
\begin{equation}
    \dot{\mathbf{x}} = \mathbf{f}(\mathbf{x}) \approx \mathbf{f}(\mathbf{x}_0) + \mathcal{M}(\mathbf{x}_0)(\mathbf{x} - \mathbf{x_0})^T + \mathcal{O}(\left|\left| \mathbf{x} - \mathbf{x}_0\right|\right|^2)\,,
\end{equation}
where $ \mathcal{M}(\mathbf{x}_0) = D\mathbf{f}(\mathbf{x}_0)$ is the stability matrix or Jacobian matrix evaluated at the critical point $\mathbf{x} = \mathbf{x_0}$. Since $\mathbf{x}_0$ is a critical point $\mathbf{f}(\mathbf{x}_0)=0$, then, near the critical point, the generally nonlinear autonomous dynamical system in Eq.(\ref{dynamical}) can be approximated by the linear system
\begin{equation}
    \dot{\mathbf{x}} \approx  \mathcal{M}(\mathbf{x}_0)(\mathbf{x} - \mathbf{x_0})^T\,. \label{linearisation}
\end{equation}
This result was actually proven by Hartman and Grobmann separately and it is known as the \textit{Hartman-Grobmann theorem} \cite{Perko2008,arrowsmith1992dynamical} or \textit{the linearization theorem}. This states that if $\mathbf{x}_0$ is a hyperbolic critical point of the nonlinear dynamical system (\ref{dynamical}) then the dynamical system and its linearization (\ref{linearisation}) are topologically equivalent, i.e. there exists a homeomorphism $H: U_{\mathbf{x}_0} \rightarrow V$, where $V$ is an open set containing the origin, mapping trajectories of Eq.(\ref{dynamical}) in $U_{\mathbf{x}_0}$ onto trajectories of Eq.(\ref{linearisation}) in $V$ and that preserves orientation in the sense that if a trajectory is directed from $p_1$ to $p_2$ in $U_{\mathbf{x}_0}$, then its
image is directed from $H(p_1)$ to $H(p_2)$ in $V$. An even more powerful theorem was proven only by Hartman and states that if $\mathbf{f}\in C^2(U_{\mathbf{x}_0})$ and the critical point is hyperbolic, then it is possible to find a $C^1$-diffeomorphism between the nonlinear dynamical system and its linearization; thus the differential structure near the critical point is preserved. The latter theorem allows us to study the linearized solution of the dynamical system (\ref{eq:variables_system4}) when the universe is near to an accelerated de Sitter expansion. Moreover, this solution should ultimately be confronted with observational data. A similar approach was used in Ref.\cite{Hrycyna2013} in the context of Brans-Dicke cosmology. Notice that if the vector function defined by the right-hand side part of the dynamical system in Eq.(\ref{eq:variables_system4}) is of class $C^{\infty}$ over a neighborhood of the critical point in Eq.(\ref{critical point}) where $k\neq 0$, then it is possible to use the linearization theorems. The linearization of the dynamical system (\ref{eq:variables_system4}) in a neighborhood of the critical point (\ref{critical point}), defining $\widetilde{X} \equiv  X -\frac{k}{3 (k+2 m-1)}$, is
\begin{equation}\left(
    \begin{array}{c}
        Z' \\ \widetilde{X}' \\ Y' \\ V'
    \end{array}
    \right)
    = \left(
    \begin{array}{cccc}
        -\frac{6 m}{k-1}-6 & 0 & \frac{9 (k+2 m-1)}{(k-1) k} & 0 \\
        0 & 0 & 1 & 0 \\
        -\frac{2 m}{k-1}+\omega -1 & \frac{9 (\omega +1) (k+2 m-1)}{k} & \frac{6 m}{k-1}-3 \omega  & 3 \omega -1 \\
        0 & 0 & 0 & -4 \\
    \end{array}
    \right)
    \left(
    \begin{array}{c}
        Z \\ \widetilde{X} \\Y \\V
    \end{array}\,\right). \label{linear system}
\end{equation}
To obtain noncomplex solutions for the Hubble parameter, in the next subsections we analyze the following two cases: $\alpha \ge 0$ and $\alpha <0$.

\subsection{Case \texorpdfstring{$\alpha \ge 0$}{}}

If $\alpha \ge 0$ the solutions of the linearization (\ref{linear system}) are given by
\begin{align}
    Z(N)\approx & b e^{-3 N (\omega +1)} +F e^{-4 N}+ c \exp \left[-\frac{3}{2}\left(1 + \frac{\sqrt{\alpha}}{k(k-1)} \right)N\right] \nonumber\\ &+d \exp \left[-\frac{3}{2}\left(1 - \frac{\sqrt{\alpha}}{k(k-1)} \right)N\right]\,, \\
    X(N) \approx &\frac{k}{3 (k+2 m-1)} \nonumber\\
    & -\frac{1}{18 (k+2 m-1)}\bigg( -\frac{2 b k e^{-3 N (\omega +1)} (k \omega -k-2 m-\omega +1)}{\omega +1} + \nonumber\\  
    & \frac{c k \left(\sqrt{\alpha}+k (-3 k-4 m+5)+4
    m-2\right) \exp \left[-\frac{3}{2}\left(1 + \frac{\sqrt{\alpha}}{k(k-1)} \right)N\right]}{k+2 m-1} \nonumber\\ 
    &  -\frac{d k
    \left(\sqrt{\alpha}+k (3 k+4 m-5)-4 m+2\right) \exp \left[-\frac{3}{2}\left(1 - \frac{\sqrt{\alpha}}{k(k-1)} \right)N\right]}{k+2 m-1} \nonumber\\ 
    & +F k e^{-4 N} (k+3 m-1)\bigg)\,,
\end{align}
\begin{align}
    Y(N) \approx & \frac{(k-1) k }{9 (k+2 m-1)}\left(\frac{3 b e^{-3 N (\omega +1)} (-k\omega +k+2 m+\omega -1)}{k-1} \right. \nonumber\\ \nonumber &\left. -\frac{3 c \left(\sqrt{\alpha}+k (-3 k-4
    m+3)\right) \exp \left[-\frac{3}{2}\left(1 + \frac{\sqrt{\alpha}}{k(k-1)} \right)N\right]}{2 (k-1) k} + \right. \nonumber\\ & \left.\frac{3 d
    \left(\sqrt{\alpha}+k (3 k+4 m-3)\right) \exp\left[-\frac{3}{2}\left(1 - \frac{\sqrt{\alpha}}{k(k-1)} \right)N\right]}{2 (k-1) k}+ \right. \nonumber\\ 
    & \left.F e^{-4 N} \left(\frac{6 m}{k-1}+2\right)\right), \\
    V(N) \equiv & \Omega_r(N) \approx \Omega_{0,r}\exp(-4N)\,, \label{radiation}
\end{align}
where $b$, $c$, $d$ are integration constants and
\begin{equation}
    F = \frac{18 \Omega_{0,r} (k+2 m-1)}{7 k^2+27 k m-16 k+18 m^2-27 m+9}\,. \label{F restriction}
\end{equation}

Notice that these solutions are only valid in a neighborhood of the critical point, i.e. when $\left|\left| \mathbf{x}-\mathbf{x}_0 \right|\right| < \varepsilon $ with $\mathbf{x} = (Z,X,Y,V)$. Based on these solutions, it is possible to compute some important cosmological quantities such as functions of $N = \ln a$. 

We can see directly from the solution (\ref{radiation}) that the evolution of the radiation density near the critical point is
\begin{equation}
    \Omega_r(N) = \Omega_{0,r}\exp(-4N)\,,
\end{equation}
where $\Omega_{0,r}$ is the radiation density today. When substituting the variable $N = \ln a$ this expression can be rewritten as
\begin{equation}
    \Omega_{r}(a) = \frac{\Omega_{0,r}}{a^4}\,,
\end{equation}
which is the standard evolution of the radiation density parameter in a de Sitter universe with respect to the scale factor. Since the solutions are in a neighborhood of the critical point, every nonlinear term in the cosmological parameters vanishes and then only linear solutions remain. 

\noindent The density parameter of the effective fluid is
\begin{equation}
\Omega_{\rm eff} = 3f_B + 2f_T - f'_B - \frac{f}{6H^2} + \left(\frac{H'}{H}\right)f_B\,,
\end{equation}
which can be written in terms of the dynamical variables as
\begin{equation}
\Omega_{\rm eff}=\frac{(3+Z)}{k}\widetilde{X}(k+2m-1)  + \frac{3+Z}{3} -Y\,,
\end{equation}
so that its linearized solution is
\begin{equation}
 \Omega_{\rm eff} \approx 1+  \frac{1}{3} b e^{-3 N (\omega +1)} \left(\frac{(k-1) k \omega }{k+2 m-1}-\frac{2 (k+m-1)}{\omega +1}\right)- \Omega_{0,r}e^{-4 N}\,.
\end{equation}
Analogously, since the first Friedmann equation reads as $\Omega_{\omega} + \Omega_{r} + \Omega_{eff} = 1$, we have
\begin{equation}
    \Omega_{\omega}\approx -\frac{1}{3} b e^{-3 N (\omega +1)} \left(\frac{(k-1) k \omega }{k+2 m-1}-\frac{2 (k+m-1)}{\omega +1}\right) \equiv \Omega_{0,\omega}e^{-3 N (\omega +1)}\,.
\end{equation}

\noindent On the other hand, we have that
\begin{align}
    \frac{H'}{H} = Z(N) \approx&  b e^{-3 N (\omega +1)}+ c \exp \left[-\frac{3}{2}\left(1 + \frac{\sqrt{\alpha}}{k(k-1)} \right)N\right]+ \nonumber\\
    &d \exp \left[-\frac{3}{2}\left(1 - \frac{\sqrt{\alpha}}{k(k-1)} \right)N\right]+F e^{-4 N}.
\end{align}
Integrating both sides of the latter equation and applying the exponential function, we find that the linearized solution of the Hubble parameter is
\begin{equation}
    \frac{H(N)}{H_i} \approx 1 -\frac{b e^{-3 N (\omega +1)}}{3 (\omega +1)}-\frac{2 c e^{-\frac{3}{2} N \left(\frac{\sqrt{\alpha }}{(k-1) k}+1\right)}}{3 \left(\frac{\sqrt{\alpha }}{(k-1)
    k}+1\right)}-\frac{2 d e^{-\frac{3}{2} N \left(\frac{\sqrt{\alpha }}{k-k^2}+1\right)}}{3 \left(\frac{\sqrt{\alpha }}{k-k^2}+1\right)}-\frac{1}{4} F e^{-4 N}\,,
\end{equation}
where $H_i$ is an integration constant. 

\noindent Finally, since the EoS of the effective fluid is
\begin{equation}
\omega_{\rm eff} = -1 + \frac{\ddot{f}_B -3H\dot{f}_B - 2\dot{H}f_T - 2H\dot{f}_T}{3H^2(3f_B + 2f_T) - 3H\dot{f}_B + 3\dot{H}f_B - \frac{1}{2}f}\,,
\end{equation}
this EoS can be written in terms of the dynamical variables as
\begin{equation}
    \omega_{\rm eff} = - 1 + \frac{Y'+ Y(Z-3) - \frac{2mXZ}{k}(3+Z) - \frac{2m}{k}\left[Y(3+Z)+X(\lambda -2Z^2)\right]}{3\frac{x}{k}(3+Z)(k+2m-1) -3Y}\,,
\end{equation}
and then, its linearized solution will be given by
\begin{align}
    \omega_{\rm eff} \approx & -1 + \nonumber\\ 
    & \left( -\frac{4 e^{-4 N} \left(18 (k+1) m^2+3 (k (2 k+7)-3) m+7 (k-1) k\right) \Omega _{0,r}}{7 k^2+27 (k-1) m-16 k+18 m^2+9} \right. \nonumber\\ 
    &\left.+\frac{b e^{-3 N (\omega +1)} \left((k-1) k (2 m+1) \omega
    -2 (k+2 m-1) (k m+k+m)+(k-1) k \omega ^2\right)}{k+2 m-1} \right. \nonumber\\ 
    & \left. +\frac{c \left(-3 k^2 m+k ((3-4 m) m-2)+\left(\sqrt{\alpha }-4\right) m+2\right) \exp \left[-\frac{3}{2}\left(1 + \frac{\sqrt{\alpha}}{k(k-1)} \right)N\right]}{k+2 m-1}\right. \nonumber\\ 
    &\left.-\frac{d \left(3 k^2 m+k (m (4
    m-3)+2)+\left(\sqrt{\alpha }+4\right) m-2\right) \exp\left[-\frac{3}{2}\left(1 - \frac{\sqrt{\alpha}}{k(k-1)} \right)N\right]}{k+2 m-1}\right)\div \nonumber\\ 
    & \Bigg(\frac{b e^{-3 N (\omega +1)} \left(-2 (k+m-1) (k+2 m-1)+(k-1) k \omega ^2+(k-1) k\omega \right)}{(\omega +1) (k+2 m-1)} \nonumber\\
    &-3 \Omega_{0,r} e^{-4 N}+3\Bigg)\,.
\end{align}

\subsection{Case \texorpdfstring{$\alpha < 0$}{}}
In this case, the solutions of the linearization in Eq.(\ref{linear system}) give real solutions on the cosmological parameters, which are
\begin{align}
    Z(N)\approx & b e^{-3 N (\omega +1)}+2 c e^{-3 N/2} \cos (N \epsilon )+F e^{-4 N}\,, \\
    X(N)\approx &\frac{b k e^{-3 N (\omega +1)} (k (\omega -1)-2 m-\omega +1)}{9 (\omega +1) (k+2 m-1)}-\frac{F k e^{-4 N} (k+3 m-1)}{18 (k+2 m-1)} \nonumber\\ 
    &+e^{-3 N/2} \left(\frac{c (k-1) k (3 k+4 m-2) \cos (\epsilon N )}{9 (k+2
    m-1)^2}-\frac{\sqrt{|\alpha| } c k \sin (\epsilon N )}{9 (k+2 m-1)^2}\right)\,, \\
    Y(N) \approx & \frac{b k e^{-3 N (\omega +1)} (-k\omega+k+2 m+\omega -1)}{3 (k+2 m-1)} +\frac{2 F k e^{-4 N} (k+3 m-1)}{9 (k+2 m-1)} \nonumber\\ 
    &+e^{-3 N/2} \left(\frac{c k (3 k+4 m-3) \cos (\epsilon N )}{3 (k+2 m-1)}-\frac{\sqrt{|\alpha| } c \sin(\epsilon N )}{3 (k+2 m-1)}\right)\,, \\
    V(N) \equiv & \Omega_r(N) \approx \Omega_{0,r}e^{-4N}\,,
\end{align}
where
\begin{equation}
    \epsilon = \frac{3 \sqrt{|\alpha | }}{2 (k-1) k}\,,\label{epsilon}
\end{equation}
and with the same restriction as Eq.(\ref{F restriction}) on $F$. 

\noindent In this case, the density parameters near the critical point are
\begin{align}
    \Omega_{\rm eff} &\approx  1+ \frac{b e^{-3 N (\omega +1)} \left(-2 (k+m-1) (k+2 m-1)+(k-1) k \omega ^2+(k-1) k \omega \right)}{3 (\omega +1) (k+2 m-1)}\label{eq:effectivedensity}\\  &\hspace{0.6cm} -\Omega_{0,r}e^{-4 N}\,, \nonumber\\
    \Omega_{\omega} &\approx -\frac{b e^{-3 N (\omega +1)} \left(-2 (k+m-1) (k+2 m-1)+(k-1) k \omega ^2+(k-1) k \omega \right)}{3 (\omega +1) (k+2 m-1)}\,,\label{eq:matteredensity}\\
    \Omega_r &\approx  \Omega_{0,r}e^{-4 N}\,, \label{eq:radiutiondensity}
\end{align}
where the Hubble parameter is then given by
\begin{equation}
    \frac{H(N)}{H_i} \approx 1-\frac{b e^{-3 N (\omega +1)}}{3 (\omega +1)}+\frac{4 c e^{-3 N/2} (2 \epsilon  \sin (N \epsilon )-3 \cos (N \epsilon ))}{4 \epsilon ^2+9}-\frac{1}{4} F e^{-4 N}\,, \label{hubble2}
\end{equation}
and the EoS of the effective fluid is
\begin{align}
    \omega_{\rm eff} &\approx -1 + \left(-\frac{4 e^{-4 N} \left(18 (k+1) m^2+3 (k (2 k+7)-3) m+7 (k-1) k\right) \Omega _{0,r}}{7 k^2+27 (k-1) m-16 k+18 m^2+9} \right. \nonumber\\ 
    & \left.+\frac{b e^{-3 N (\omega +1)} \left((k-1) k \omega  (k-2
    m-1)-2 ((k-1) k-m) (k+2 m-1)+(k-1)^2 k \omega ^2\right)}{(k-1) (k+2 m-1)}\right. \nonumber\\ 
    & \left.+e^{-3 N/2} \left(\frac{2 \sqrt{|\alpha| } c m \sin (N \epsilon )}{k+2 m-1}-\frac{2 c (k (m (3 k+4
    m-3)+2)+4 m-2) \cos (N \epsilon )}{k+2 m-1}\right)\right) \div \nonumber\\ 
    & \Bigg( \frac{b e^{-3 N (\omega +1)} \left(-2 (k+m-1) (k+2 m-1)+(k-1) k \omega ^2+(k-1) k \omega \right)}{(\omega +1) (k+2 m-1)} \nonumber\\
    &-3 \Omega_{0,r} e^{-4 N}+3\Bigg)\,. \label{eq:EoSsecodnbranch}
\end{align}
We have to keep in mind that the solutions described for both branches $\alpha \ge0$ and $\alpha <0$ are only valid in a neighborhood of the critical point, i.e., they describe the dynamics of a universe close to being a perfect de Sitter universe, which in terms of the parameter $N= \ln a$, corresponds to $N \to \infty$, and in terms of the redshift, corresponds to $z \to -1$. The theory of dynamical systems cannot guarantee these solutions to work well for a wide range of redshift. However, as we will see, when confronting observational data, we will notice that some models of $f(T,B)$ can describe the dynamics of the Universe today and may give a glimpse into the solution to the $H_0$ tension.

\section{Phenomenological hints on \texorpdfstring{$H_0$}{} tension from the generic \texorpdfstring{$f(T,B)$}{} dynamical system}
\label{sec:h0tension}

As we discussed, the theory of dynamical systems guarantees that the equations obtained above for the cosmological parameters are a solution of the Friedmann equations when $N \to \infty$, or $z \to -1$ in terms of the redshift. Therefore, by studying this solution in phenomenological scenarios where cosmological data is available, we can better test this solution.

In this section, we will use our results derived from our nonlinear dynamical system and link them with current $H(z)$ observational data to investigate in which redshift range this system of dynamical equations still holds as a valid solution for the Friedmann equations.

As a first step in that direction, we consider in this proposal Hubble parameter measurements $H(z)$ \cite{Magaa2018}, in which the current sample consists of 51 measurements in the redshift range $0.07< z < 2.0$ of which 31 points correspond to cosmic chronometers and the rest are from baryonic acoustic oscillation (BAO) estimates from late times. Afterwards, we will constrain the values for $\Omega_{0, r} $, $ k $, $ m $, $ \omega $, and the integration constants, and also perform this for the $\alpha <0$ case to ensure that the obtained values of $k$ and $m$ provide a stable critical point.

\begin{figure}[H]
\begin{center}
\includegraphics[scale=0.7]{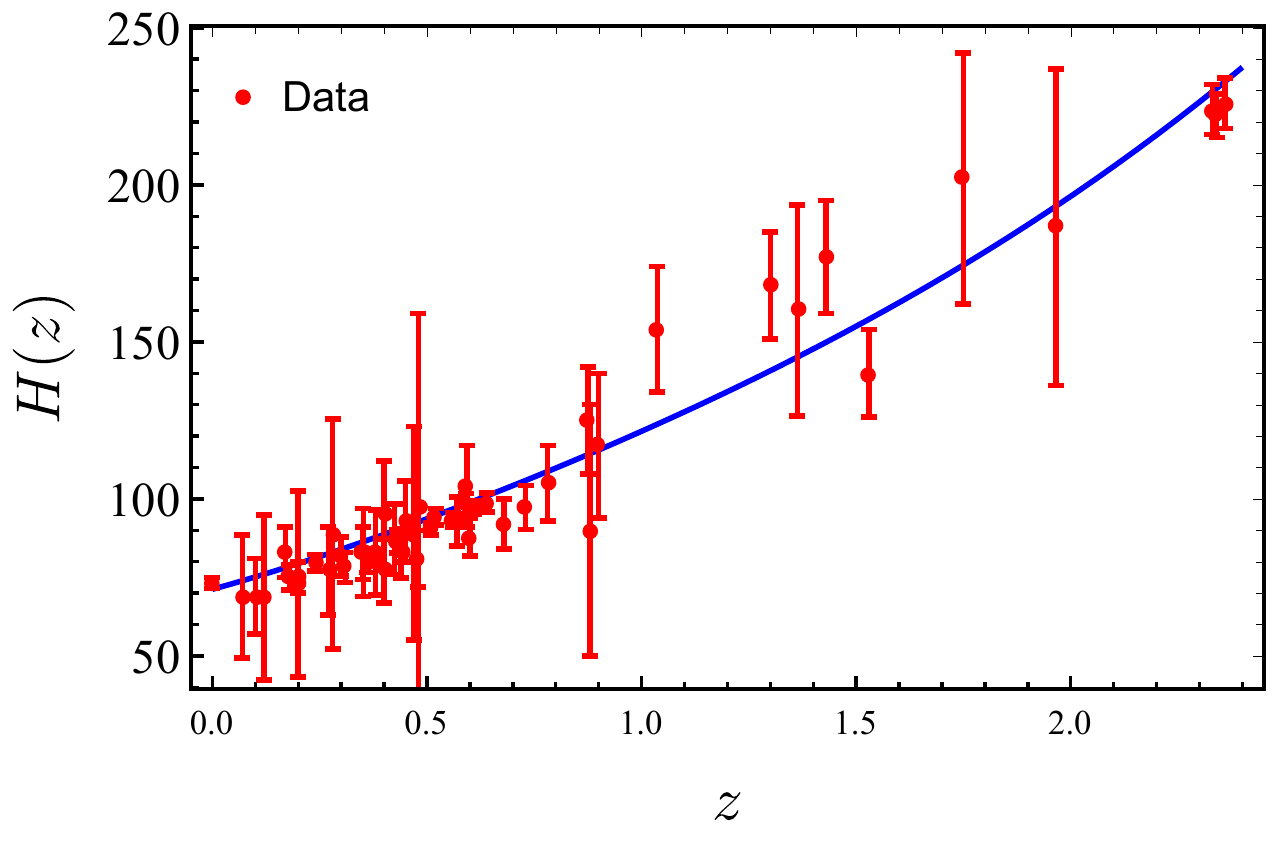}
\caption{Reconstruction of the evolution of $H(z)$ using the best fit obtained with the model \protect (\ref{hubble2}) (blue solid line) in comparison to the observational data described  \cite{Magaa2018,Escamilla-Rivera:2020giy} (red dots).}\label{hubblewerrors}
\end{center}
\end{figure}

To obtain a convenient set of free parameters appearing in Eq.(\ref{hubble2}) we must perform the change of variable $N = \ln \frac{1}{1+z}$, in order to obtain an expression in terms of the redshift and then fit this equation to the available data of the $H(z)$ \cite{Magaa2018} with respect to the redshift. If we consider the free parameters without any restrictions, a nonphysical set of parameters may be obtained from the fitting, like a negative value for the mean value of the radiation density $\Omega_{0,r}$ or an EoS for matter less than $-1$ which leads to an instability, to name a few, then a more mindful approach of fitting has to be performed. To obtain a convenient set of parameters with physical relevance, we have to restrict or associate certain values to these parameters. More to the point, it has been shown that the radiation density today has a small value of $\Omega_{0,r}\approx 10^{-5}$ \cite{Lahav:2019bbc}; then in order to simplify the model, we consider $\Omega_{0,r}=0$. We restrict, based upon observations in Refs.\cite{Planck2018,Abbott2020,Suyu2017}, $H_0$ to vary in the range of $60 \le H_0 \le 90$. Finally, we consider the case $\alpha <0$ for this analysis since any values of $k$ and $m$ in this case provide a stable critical point. Under these considerations, it is possible to avoid fitting $k$ and $m$ directly since it may lead to problems in the numerical code, due to the square root in Eq.(\ref{hubble2}), but it is possible and feasible to fit $\epsilon$ and then, based on Eq.(\ref{epsilon}), obtain values of $k$ and $m$ in the level curve of the obtained value for $\epsilon$. Based on these considerations, the best-fit parameters are
\begin{align}
    \omega &= 0.000 \pm 2.951\,,\quad \quad  \epsilon = 1.749 \pm 5.671\,,  \quad b =  -0.200 \pm 2.735\,, \nonumber\\ 
       c &= -0.213 \pm 2.221\,,\quad  H_i = 60.000 \pm 10.973\,, \label{fittedvalues}
\end{align}
and the possible set of values of $k$ and $m$ in the level curve $\epsilon(k,m)= 1.749$ are
\begin{equation}
    k \approx -0.294\,, \quad m = 0.75\,.
\end{equation}
If we evaluate the best fit at $z=0$ we obtain that $H_0 \approx 71.236 \pm 144.229$. In Fig.~\ref{hubblewerrors} the evolution of the approximated solution for the Hubble parameter (\ref{hubble2}) with the fitted values (\ref{fittedvalues}) and the observational data from Ref.\cite{Magaa2018} is shown.

The method described above leads to high error regions (C.L) in the fitted parameters and a high error in the $H_0$ value due to the large number of free parameters and the small amount of observational data. A different way to link the observational data with the approximated solutions obtained for the dynamical system can be performed. By taking the stability analysis performed in Sec. \ref{sec:dynamicalsec}, we can choose some values for $k$ and $m$ from the analysis and reduce the number of free parameters in the fitting procedure, but we still obtain a high error C.L are obtained for the remaining free parameters. In order to further reduce this propagation of errors, we set some of the free parameters to specific values, such as $\omega = 0$ (which corresponds to a dust model and reduces the error C.L to the remaining parameters and the $H_0$ value).

\begin{figure}[H]
\begin{center}
\includegraphics[scale=0.6]{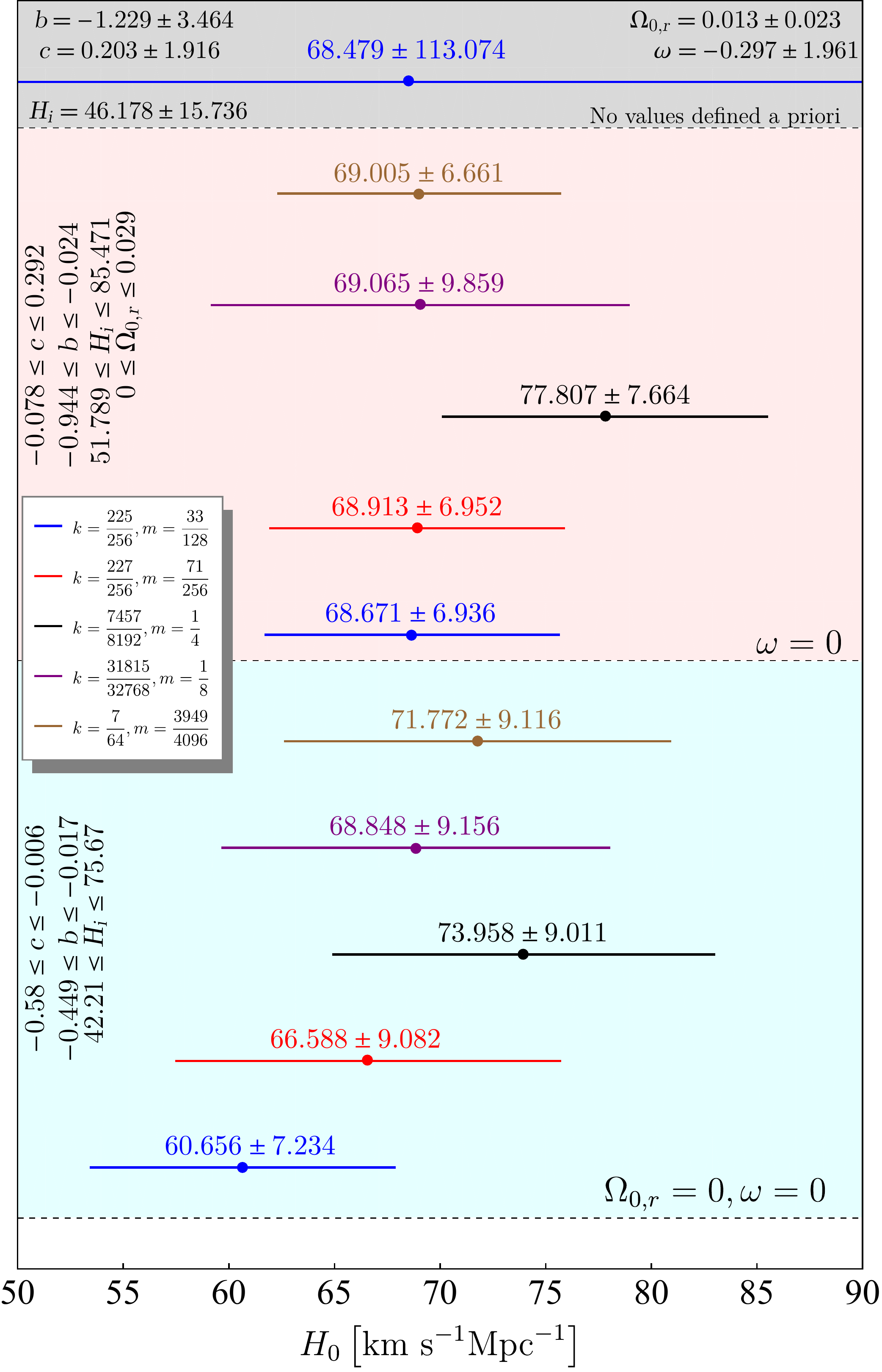}
\caption{Values obtained for $H_0$ with specific choices of $k$ and $m$ denoted in the inner box label. Top $H_0$ best-fit value for a dynamical system with all variables allowed to be free. Middle: $H_0$ best-fit values for a dust model $\omega =0$. Bottom: $H_0$ best-fit values for a dust and zero radiation model.}
\label{fig:hubbletension}
\end{center}
\end{figure}

In Fig.(\ref{fig:hubbletension}) we show a plot with different values of $H_0$ obtained under different conditions for some values of $k$ and $m$ obtained from the dynamical analysis. Describing the regions denoted by colors, we have that: 
the cyan region shows different values of $H_0$ for a dust $\omega=0$ and zero radiation $\Omega_{0,r}=0$ model in which the fitted parameters satisfy $-0.580 \le c \le -0.006 $, $-0.449 \le b \le -0.017$ and $42.21 \le H_i \le 75.670$.

The pink region shows different values for $H_0$ for a dust model with radiation; in this case the fitted parameters satisfy $-0.078 \le c \le 0.292$, $-0.944 \le b \le -0.024$, $51.789 \le H_i \le 85.471$  and $0 \le \Omega_{0,r} \le 0.029$.

In the gray region only one value of $H_0$ is shown, since no values are defined \textit{a priori}, the error C.L becomes large, and only one value is plotted to show explicitly the large error in this case. The obtained value for $H_0$ corresponds to the fitted parameters $b = -1.229 \pm 3.464$, $c=0.203 \pm 1.916$, $H_i = 46.178 \pm 15.736$, $\Omega_{0,r}=0.013 \pm 0.023$ and $\omega = -0.297 \pm 1.961$.

\begin{figure}[H]
\centering
\includegraphics[scale=0.5]{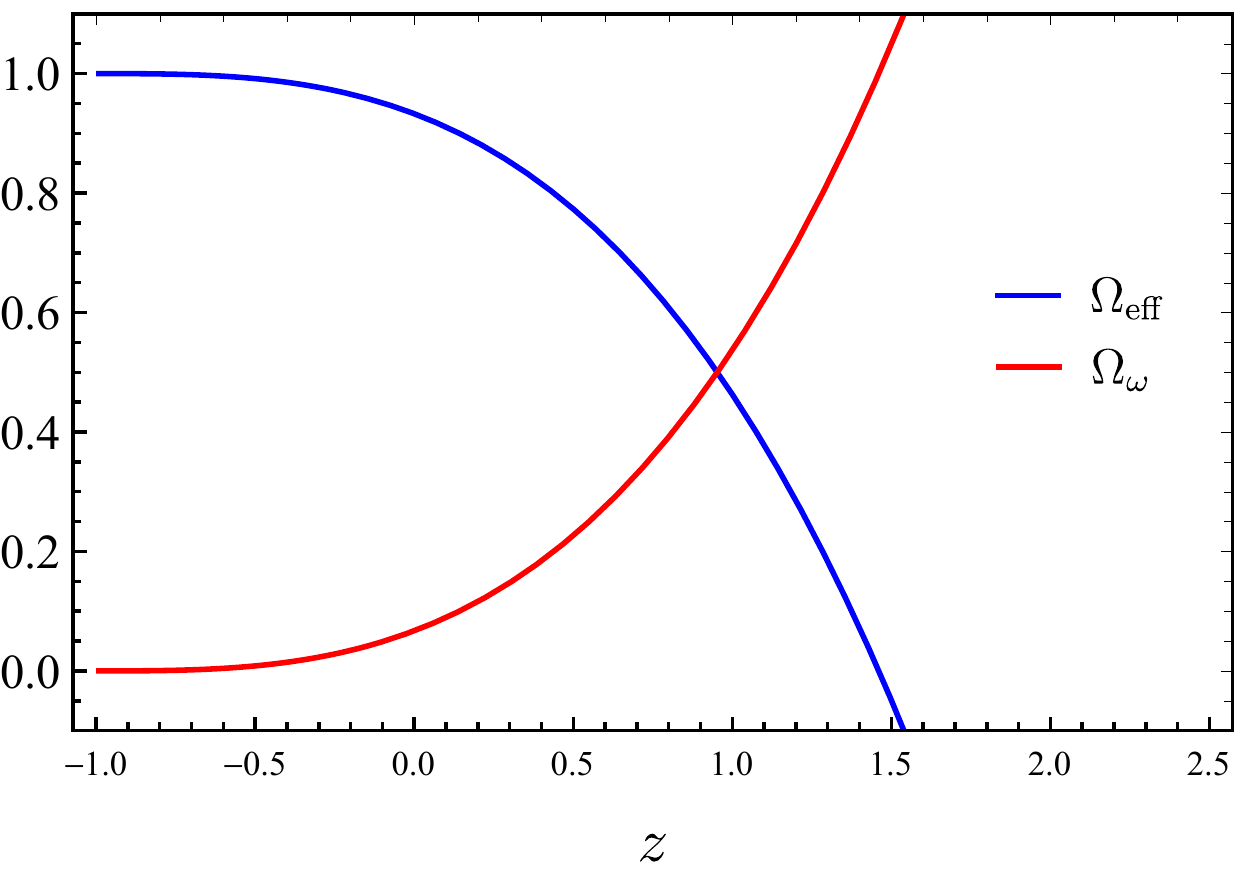}
\includegraphics[scale=0.55]{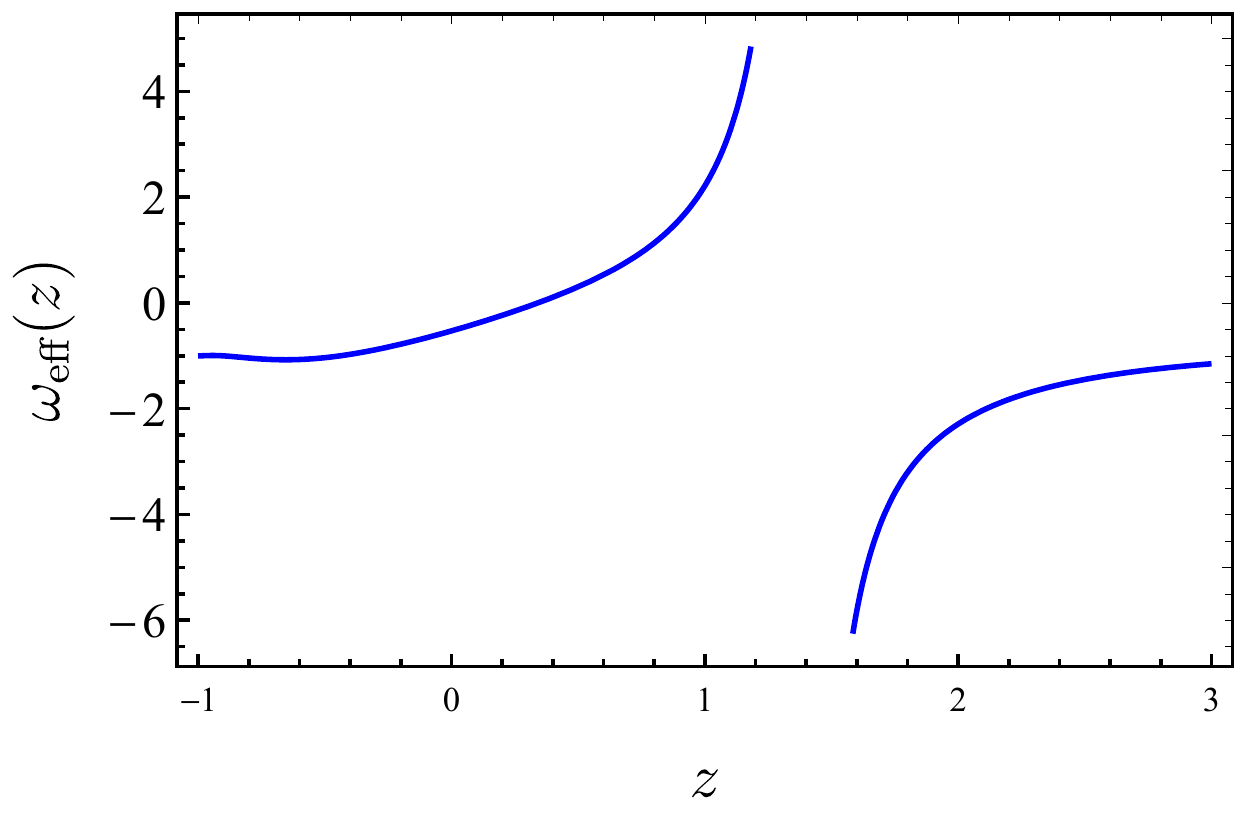}
\caption{Left: evolution of the density parameters given by Eqs. (\ref{eq:effectivedensity}) and (\ref{eq:matteredensity}) in terms of the redshift $z$. Right: evolution of $\omega_{eff} (z)$ given by Eq.(\ref{eq:EoSsecodnbranch}). The inflection point occurs at $z\approx 1.397$ for this model.}
\label{densityparameters}
\end{figure}

With the results obtained it is also possible to plot, e.g. the density parameters, the EoS of the effective fluid and deceleration parameter with respect to the redshift, making the change of variable $N = \ln \frac{1}{1+z}$, e.g. with the fitted values (\ref{fittedvalues}) and null radiation, the graph of these cosmological parameters are shown in Fig.~\ref{densityparameters} for the density parameters (left) and for the EoS of the effective fluid (right).

\section{Conclusions}
\label{sec:conclusions}

As we have explored, a number of $f(T,B)$ models have been fitted using cosmic chronometers, supernova and baryonic acoustic oscillation data which resulted in model parameter constraints, that in some cases can alleviate the $H_0$ tension which has become very topical in recent years. Due to the complexity of the evolution of the cosmological system, some viable $f(T,B)$ models can be analyzed via their dynamical systems. One of these results was the mixed power-law model, where the torsion scalar and boundary term are coupled with independent indices. This is an important matter since this model represents a novel deviation from $f(\lc{R})$ gravity in that it cannot be reproduced by any choice of $f(\lc{R})$ model. Moreover, due to the complexity of the ensuing dynamical system, one of the dynamical variables had been set to be a constant which overly constrained the system. With the aim to find a generic formulation of this dynamical system, in this work we have explored an approach in which this assumption does not need to be taken and the fully autonomous dynamical system can be explored using their hyperbolic critical points. We used the Hartman-Grobmann theorem in order to study our nonlineal dynamical system (\ref{eq:variables_system4}) whose solution is near to an accelerated de Sitter expansion. One of the most interesting features presented in this work is the possibility to describe the dynamics of the Universe today using a narrow branch of our dynamical system, and in this case to explore the nature of the $H_0$ tension from a autonomous system point of view.

To achieve the last goal, we employed in this work a $H(z)$ observational sample from cosmic chronometers plus BAO estimates. Our purpose was to constrain the characteristic values which define the mixed power law only. According to this phenomenological analysis, it was possible to find an optimal set of cases where the $H_0$ tension can be alleviated using specific values for the mixed $f(T,B)$ power law as we can see from Fig.~\ref{fig:hubbletension}. In these analyses, we considered three scenarios: (1) all the parameters allowed to be free; (2) a dust model, and; (3) a dust solution with zero radiation. According to this, cases where $k < 0.9$ seem to be in agreement with the \textit{Planck} 2018 $H_0$ constraint at $1\sigma$, while for the specific case $k=7457/8192$ and $1/4$, the $H_0$ is quite high with a considerable error propagation. This novel feature obtained from a specific mixed $f(T,B)$ model cannot be derived from a Bayesian analysis directly (see Ref.\cite{Escamilla-Rivera:2019ulu} for instance), which provided an interesting way to relax the condition $\lambda=\text{const}$ where linearized solutions can be found in order to fix a set of points that solves the $H_0$ tension. In conclusion, the mixed power-law model analyzed here is novel in that it is not reproduced by an $f(\lc{R})$ gravity Lagrangian, and it been shown promising in reproducing a cosmology that resonates with current observations, giving motivation to further studies in this direction.

\section*{Acknowledgements}
\label{sec:acknowledgements} 
CE-R is supported by the Royal Astronomical Society as FRAS 10147. CE-R and GAR-F are supported by PAPIIT Project IA100220. This article is based upon work from CANTATA COST (European Cooperation in Science and Technology) action CA15117, EU Framework Programme Horizon 2020. The authors would like to acknowledge networking support by the COST Action CA18108 and funding support from Cosmology@MALTA which is supported by the University of Malta.

\bibliographystyle{utphys}
\bibliography{refs2}

\end{document}